\documentclass[twocolumn]{aastex62}

\shortauthors{Hou et al.}

\begin{document}

\title{Magnetic Evolution of Highly-Sheared Region in Active Region 13842 Producing Large X9.0 Flare}

\correspondingauthor{Yijun Hou}
\email{yijunhou@nao.cas.cn}

\author[0000-0002-9534-1638]{Yijun Hou}
\affiliation{State Key Laboratory of Solar Activity and Space Weather, National Astronomical Observatories, Chinese Academy of Science, Beijing 100101, China}
\affiliation{School of Astronomy and Space Science, University of Chinese Academy of Sciences, Beijing 100049, China}
\affiliation{Yunnan Key Laboratory of the Solar physics and Space Science, Kunming 650216, China}

\author[0000-0001-6655-1743]{Ting Li}
\affiliation{State Key Laboratory of Solar Activity and Space Weather, National Space Science Center, Chinese Academy of Sciences, Beijing 100190, China}
\affiliation{School of Astronomy and Space Science, University of Chinese Academy of Sciences, Beijing 100049, China}
\affiliation{State Key Laboratory of Solar Activity and Space Weather, National Astronomical Observatories, Chinese Academy of Science, Beijing 100101, China}

\author[0000-0002-6565-3251]{Shuhong Yang}
\affiliation{State Key Laboratory of Solar Activity and Space Weather, National Astronomical Observatories, Chinese Academy of Science, Beijing 100101, China}
\affiliation{School of Astronomy and Space Science, University of Chinese Academy of Sciences, Beijing 100049, China}

\author[0000-0001-5776-056X]{Leping Li}
\affiliation{State Key Laboratory of Solar Activity and Space Weather, National Astronomical Observatories, Chinese Academy of Science, Beijing 100101, China}

\author[0009-0007-4469-0663]{Yingjie Cai}
\affiliation{State Key Laboratory of Solar Activity and Space Weather, National Astronomical Observatories, Chinese Academy of Science, Beijing 100101, China}
\affiliation{School of Astronomy and Space Science, University of Chinese Academy of Sciences, Beijing 100049, China}

\author[0000-0002-3657-3172]{Xiaofeng Liu}
\affiliation{Purple Mountain Observatory, Chinese Academy of Sciences, Nanjing 210034, China}
\affiliation{School of Astronomy and Space Science, University of Science and Technology of China, Hefei 230026, China}
\affiliation{State Key Laboratory of Solar Activity and Space Weather, National Astronomical Observatories, Chinese Academy of Science, Beijing 100101, China}

\author{Shuo Yang}
\affiliation{North China University of Technology, Beijing 100144, China}

\author[0000-0001-9893-1281]{Yilin Guo}
\affiliation{Beijing Planetarium, Beijing Academy of Science and Technology, Beijing 100044, China}

\author[0000-0002-7544-6926]{Shihao Rao}
\affiliation{School of Astronomy and Space Science, Nanjing University, Nanjing 210023, China}
\affiliation{Key Laboratory of Modern Astronomy and Astrophysics (Nanjing University), Ministry of Education, Nanjing 210023, China}

\author[0000-0001-7693-4908]{Chuan Li}
\affiliation{School of Astronomy and Space Science, Nanjing University, Nanjing 210023, China}
\affiliation{Key Laboratory of Modern Astronomy and Astrophysics (Nanjing University), Ministry of Education, Nanjing 210023, China}

\author{Guiping Zhou}
\affiliation{State Key Laboratory of Solar Activity and Space Weather, National Astronomical Observatories, Chinese Academy of Science, Beijing 100101, China}
\affiliation{School of Astronomy and Space Science, University of Chinese Academy of Sciences, Beijing 100049, China}

\begin{abstract}
Shearing motion and magnetic flux cancellation around the polarity inversion line (PIL) play significant roles in the build-up of free magnetic
energy and magnetic flux rope (MFR) in source region of major solar flares. Here we investigate the magnetic evolution of a highly-sheared PIL
in active region (AR) 13842, hosting the largest X9.0 flare of Solar Cycle 25. Since 2024 September 29, a positive-polarity pore persistently
drifted northward along the western side of the AR's main negative-polarity sunspot. The main sunspot remained stationary until negative-polarity
patches successively emerged to its east and approached. Rear-ended by these same-polarity patches, the sunspot then began moving westward toward
the opposite-polarity pore around October 1, forming a collisional PIL. Meanwhile, on the PIL's other side, the pore was also rear-ended by
same-polarity patches sequentially emerging behind it, accelerating the shearing motion around the PIL, where frequent flux cancellations were
also observed. Synchronous rapid accumulation of free magnetic energy and formation of MFR were then observed in the PIL, where multiple major
flares successively occurred within two days. Before these large flares, the area and total free energy of the high-free-energy-density PIL region
gradually decreased in the photosphere, which could be caused by the initial ascent of MFR before eruption and serve as a precursor of solar eruptions.
These results suggest that persistent flux emergences with cross separation directions facilitates rapid formation of collisional shearing PIL and
frequent flux cancellations, leading to repeated MFR formations and multiple large flares in a relatively short time.
\end{abstract}



\section{Introduction} \label{sect1}
Solar flares are the most energetic phenomena in our solar system and manifest as abrupt increase in electromagnetic emission of Sun's local region over a
wide spectral range such as $\gamma$-ray, X-rays, (E)UV, radio, and even in white light \citep{1999ApJ...512..454D,2011SSRv..159...19F,2023ApJ...959...69H,
2024ApJ...975...69C,2026RAA....26d7001C}. During solar flares, dramatic amounts of free energy stored in the magnetic fields of solar atmosphere
are suddenly released and converted into kinetic energy of accelerated particles and thermal energy, most probably through magnetic reconnection
\citep{2002AARv..10..313P,2011LRSP....8....6S,2014ApJ...797L..14T,2020A&A...640A.101H}. In such ``energy storage-release" paradigm, how free magnetic
energy (i.e., energy above the potential magnetic field) builds up in pre-flare coronal magnetic fields of local flaring region is one of the most critical
aspects \citep{2019LRSP...16....3T,2023ApJ...942...27L}.

Previous studies found that major solar flares tend to take place in the vicinity of the polarity inversion lines (PILs), where the horizontal
gradient of the vertical field is steep, the transverse fields are highly sheared, and the free energy density are strongly enhanced
\citep{1973SoPh...32..173Z,1996ApJ...456..861W,2007ApJ...655L.117S,2012ApJ...748...77S,2017ApJ...834...56T,2023ApJ...950L...3C,2024ApJ...964..159L}.
And solar active regions (ARs) with compact shearing PILs are also known to be often flare-productive \citep{2019LRSP...16....3T}, such as AR 12673
\citep{2017ApJ...849L..21Y,2018ApJ...856...79Y,2018A&A...619A.100H,2018ApJ...869...13J,2020ApJ...894...29B} in Solar Cycle 24 and AR 13664
\citep{2024ApJ...972L...1L,2024ApJ...976L..12J} in Solar Cycle 25. \citet{2017ApJ...849L..21Y} proposed that in AR 12673, movement of emerging patches was
blocked by the pre-existing main sunspot and consequently distorted, forming a flare-productive PIL where a great deal of free energy was accumulated.
\citet{2018A&A...619A.100H} found that successive interactions between the different emerging dipoles and strong shearing motions around the
formed PIL also resulted in a complex system consisting of multiple magnetic flux ropes (MFRs). \citet{2019ApJ...871...67C} further suggested
that ``collisional shearing" between non-conjugated polarities of two emerging bipoles within the same AR could drive continuous magnetic flux
cancellation, accompanied by a series of solar eruptions.

At the end of 2024 September, AR 13842 appeared at the east solar limb and then rapidly evolved into a flare-productive AR with a highly-sheared
PIL, where producing two X-class (X7.1 and X9.0) and more than ten M-class flares within just 3 days from October 1 to October 3. The X9.0 flare
on October 3 is so far the largest flare of Solar Cycle 25\citep{2025ApJ...985L..16D}, which is also the largest flare in recent 7 years after X9.3
flare on 2017 September 6. This AR may raise interest from the community, but the following questions still remain to be answered: (1) What are the
magnetic characteristics of the source region of major flares produced in AR 13842? (2) Why does the free energy build up so fast in the local
flaring region of AR 13842 that the major flares could occur in the same region in such a relatively short time? In this study, we are dedicated
to answering these two questions.

\section{Observations and Data Analysis}\label{sect2}
On 2024 October 3, an X9.0 flare, the largest flare in Solar Cycle 25 so far, occurred in NOAA AR 13842, where another X7.1 flare had already happened
just about 38 hours ago on October 1. Both flares were well observed by the \emph{Solar Dynamics Observatory} \citep[\emph{SDO};][]{2012SoPh..275....3P}
and \emph{Chinese H$\alpha$ Solar Explorer} \citep[\emph{CHASE};][]{2022SCPMA..6589602L}. The Helioseismic and Magnetic Imager \citep[HMI;][]{2012SoPh..275..229S}
and Atmospheric Imaging Assembly \citep[AIA;][]{2012SoPh..275...17L} on board the \emph{SDO} successively provide full-disk one-arcsecond resolution line
of sight (LOS) magnetograms and continuum intensity maps every 45 s, as well as multi-wavelength (E)UV images with a cadence of (12)24 s and a spatial
resolution of 1.{\arcsec}2, respectively. The H$\alpha$ Imaging Spectrograph on board \emph{CHASE} can provide full-disk images of the Sun in H$\alpha$
(6559.7--6565.9 {\AA}) waveband with a cadence of about 70 s and a spatial resolution of 1.{\arcsec}2. Here we used the HMI magnetic field observations,
continuum intensity maps, AIA 94 {\AA}, 171 {\AA}, 304 {\AA}, and HIS H$\alpha$ images for investigating the evolution of flare source region in AR 13842
from September 30 to October 3. The data used for evolution of the AR's magnetic field were differentially rotated to a middle reference time of 00:00 UT
on October 2 while the data analyzed for the X9.0 and X7.1 flares were respectively aligned to the reference time of their peak times: 12:18 UT on
October 3 and 22:20 UT on October 1. It is worth noting that most HMI observations between 12:40 UT and 19:00 UT on October 2 are not available due to
the telescope equipment malfunction. The observations from the \emph{Geostationary Operational Environmental Satellite} (\emph{GOES}) were also used
to present the variation of soft X-ray (SXR) 1--8 {\AA} flux.

To reconstruct the three-dimensional (3D) magnetic configuration above the region of interest, we also utilized the ``weighted optimization" method to
perform nonlinear force-free field (NLFFF) extrapolation \citep{2004SoPh..219...87W,2012SoPh..281...37W} based on the photospheric vector magnetic
fields observed by \emph{SDO}/HMI with a cadence of 12 minutes. The NLFFF calculations were performed within a box of 648$\times$600$\times$256 uniform
grid points (235$\times$218$\times$93 Mm$^{3}$, covering the entire AR 13842) at 12:00 UT on September 30, 21:48 UT on October 1, and 12:00 UT on
October 3, respectively. Furthermore, through the method developed by \citet{2016ApJ...818..148L}, we calculated the twist number $T_{w}$ of the
extrapolated 3D magnetic fields, which represents how many turns two field lines wind about each other and plays an important role in identifying a
twisted MFR.

\section{Results and Discussion}\label{sect3}
\begin{figure*}
\centering
\includegraphics [width=0.95\textwidth]{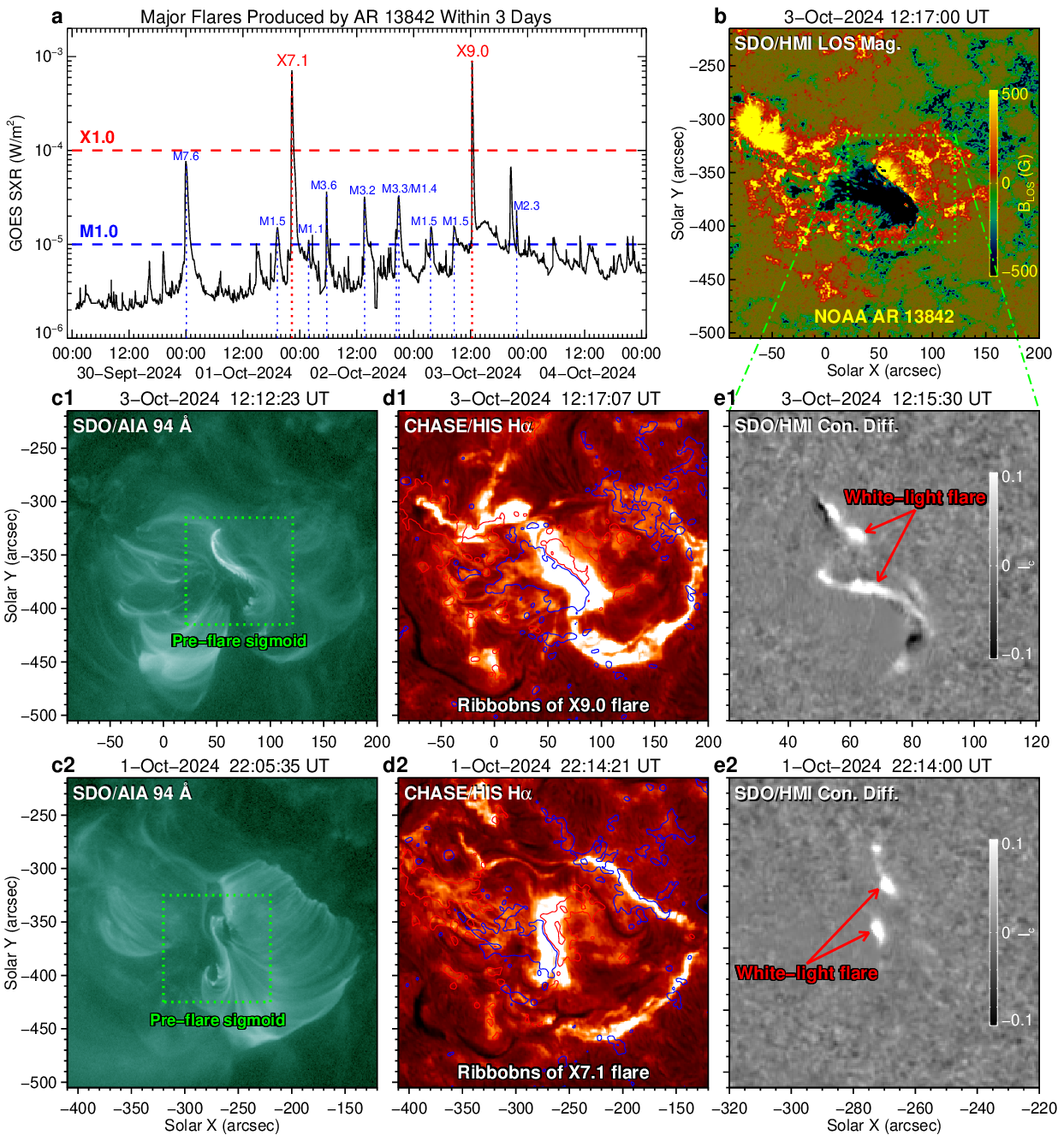}
\caption{Major flares produced by AR 13842 from 2024 October 1 to October 3.
(a): GOES SXR 1--8 {\AA} flux variation. The blue and red vertical dotted lines mark the peak time of major flares produced by AR 13842.
(b): \emph{SDO}/HMI LOS magnetogram displaying photospheric magnetic field of AR 13842. The green square outlines the field of view (FOV)
of (e1) and (e2).
(c1), (d1), and (e1): \emph{SDO}/AIA 94 {\AA} image, \emph{CHASE}/HIS H$\alpha$ image, and difference image of \emph{SDO}/HMI continuum
intensity showing the pre-flare hot sigmoid structure appearing in the flaring region of X9.0 flare on October 3 and flare ribbons
(especially the white-light ones) around its peak time. The red and blue curves overlaid in (d1) are the LOS magnetic field contours
at $\pm$ 200 G.
(c2), (d2), and (e2): Similar to (c1)--(e1), but for the X7.1 flare on October 01.
Two animations (Figure1a.mp4 and Figure1b.mp4) covering 21:40 UT to 22:40 UT on October 1 and 11:50 UT to 12:50 UT on October 3
are available online, which present the X7.1 and X9.0 flares through AIA 94 {\AA}, 171 {\AA}, 303 {\AA}, and HMI continuum intensity images.
Both animation's durations are 8 seconds.
}
\label{fig1}
\end{figure*}

\begin{figure*}
\centering
\includegraphics [width=0.95\textwidth]{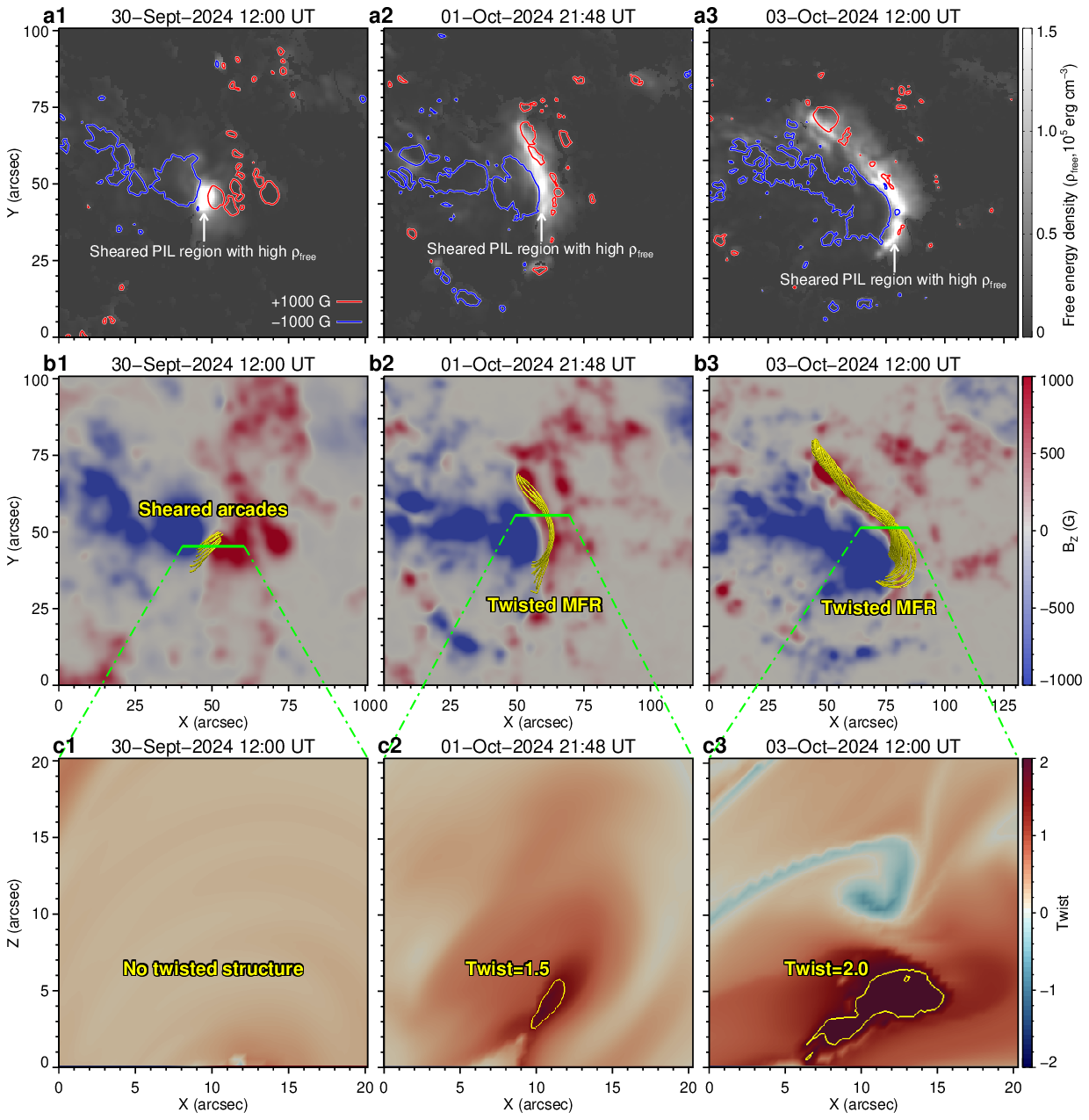}
\caption{Evolution of photospheric nonpotentiality and three-dimensional (3D) magnetic topology in flaring PIL region of AR 13842.
(a1)--(a3): Sequence of free energy density ($\rho_{free}$) maps overlaid with the vertical magnetic field contours at $\pm$ 1000 G
showing the PIL region with high $\rho_{free}$ at 12:00 UT on September 30, 21:48 UT on October 1, and 12:00 UT on October 3.
(b1)--(b3): Top view of sheared arcades and twisted MFRs above the sheared PIL region at the corresponding times.
(c1)--(c3): Distribution of magnetic twist $T_{w}$ in the vertical planes based on the green cuts labeled in (b1)--(b3). The yellow
curves in (c2) and (c3) mark $T_{w}$ contours at 1.5 and 2.0, outlining the general section shape of MFRs appearing before onset
of the X7.1 and X9.0 flares, respectively.
}
\label{fig2}
\end{figure*}

\begin{figure*}
\centering
\includegraphics [width=0.95\textwidth]{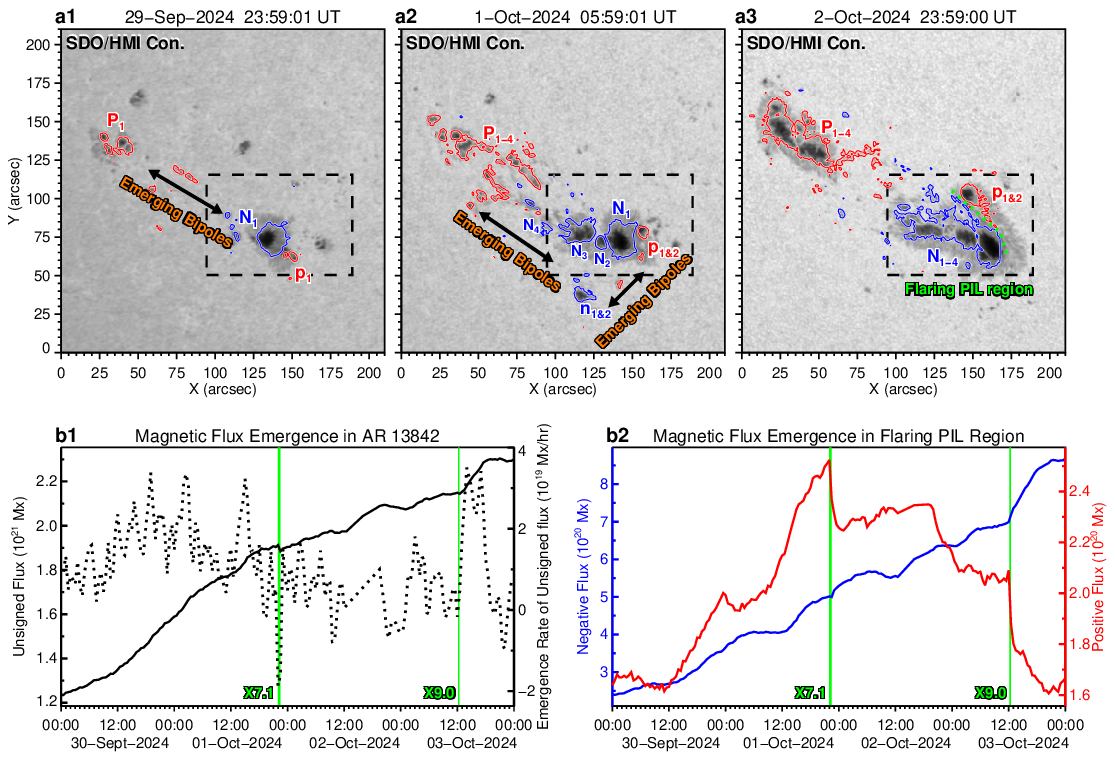}
\caption{Rapid magnetic flux emergence with cross separation directions in AR 13842.
(a1)--(a3): Sequence of HMI continuum intensity maps overlaid with the LOS magnetic field contours (red and blue curves) showing the magnetic
evolution of AR 13842 from September 30 to October 3. P$_{1}$, N$_{1}$, p$_{1}$, and n$_{1}$ label the main conjugated magnetic polarity pair of
AR 13842, and the satellite positive pore and its conjugated negative polarity, respectively. P$_{2}$, P$_{3}$, P$_{4}$, N$_{2}$, N$_{3}$, N$_{4}$,
p$_{2}$, and n$_{2}$ mark the subsequent emerging polarities during the two bipoles' evolution. Black solid arrows indicate separation directions of
the two emerging bipoles perpendicular to each other. The black rectangles outline the evolving region of the two polarities around the flaring PIL region.
(b1): Temporal evolution of the unsigned magnetic flux (solid curves) and its emergence rate (dotted curves) in the whole AR 13664.
(b2): Temporal evolution of the negative (blue curves) and positive (red curves) magnetic flux around the flaring PIL region.
An animation (Figure3.mp4) covering September 30 to October 3 is available online, which displays the evolution of spots and underlying magnetic
fields in the AR 13842. The animation's duration is 8 seconds.
}
\label{fig3}
\end{figure*}

After appearing at the east solar limb around 2024 September 28, AR 13842 rapidly developed and experienced a significant magnetic flux emergence on
September 30 between the two main sunspots (see the online animation associated with Figure \ref{fig3}). Between the AR's leading negative sunspot and
a satellite positive pore rotating around it, a highly-sheared PIL then gradually formed from September 30 to October 3, where a series of flares occurred
(see Figures \ref{fig1}(a) and (b)). The two largest flares among are X7.1 flare on October 1 (peaked at 22:20 UT) and X9.0 flare on October 3 (peaked at
12:18 UT) occurring within just 38 hours. Multi-wavelength observations of these two flares are displayed in Figures \ref{fig1}(c1)--(e1) and (c2)--(e2),
and the associated online animations. It is obvious that hot sigmoidal structures were already located in the sheared PIL region of both flares before
their onsets (see green dotted squares in Figures \ref{fig1}(c1) and (c2)), which are identified as pre-flare MFRs in recent studies \citep{2008ApJ...680..734J,
2012ApJ...746...17G, 2013ApJ...769L..25C,2016A&A...592A.138H,2024ApJ...967..130L}. Around the GOES peak of flare, both flares showed multiple flare ribbons
in H$\alpha$ observations, including two typical main ribbons located on both sides of the PIL, a larger quasi-circular ribbon in the west, and remote
brightenings in the northeast (see Figures \ref{fig1}(d1) and (d2)). Such morphology of flare ribbons imply a complex overlying magnetic environment above
the core flaring region, where may exist a large-scale fan-spine topology \citep{2013ApJ...778..139S,2019ApJ...871....4H,2019ApJ...885L..11S,
2020ApJ...898..101Y,2020ApJ...899...34L,2020A&A...636L..11Z}. In addition, HMI continuum intensity maps revealed remarkable white-light emission enhancement
signals in the main ribbons of both flares (see Figures \ref{fig1}(e1) and (e2)).

At the early evolution stage of the flare-productive PIL, the region between the main negative sunspot and the satellite pore with opposite polarity
already had a high free energy density ($\rho_{free}$) at 12:00 UT on September 30 (see bright region in Figure \ref{fig2} (a1)). The NLFFF
extrapolation result in Figure \ref{fig2} (b1) shows that there were sheared arcades connecting the main and the satellite spots above the PIL. Along
the cut marked by the green line in the sheared PIL region, we then examined the distribution of $T_{w}$ in the vertical plane, which reveals no
twisted magnetic structure at that time (panel (c1)). However, as shown in Figures \ref{fig2} (a2), (b2), and (c2), at 21:48 UT on October 1, just
before the onset of X7.1 flare, an MFR with a maximum $T_{w}$ of $\sim2.0$ formed above the PIL region, where the high $\rho_{free}$ area had became
curved and much longer due to the shearing motion between the main sunspot and the satellite pore. Figures \ref{fig2} (a3), (b3), and (c3) display
similar situations before the beginning of X9.0 flare at 12:00 UT on October 3, when the sheared PIL became longer and had larger $\rho_{free}$.
The twisted MFR above the PIL had larger $T_{w}$ with the maximum value of $\sim2.7$.

The above results reveal that during the evolution of the AR's leading main sunspot, twisted MFRs rapidly formed above the PIL region between the
sunspot and the surrounding satellite pore while the PIL region had larger and larger curvature, length, and $\rho_{free}$. Similar characteristics
were also previously reported in other flare-productive ARs like AR 12673 \citep{2018A&A...619A.100H,2020ApJ...894...29B}. It's worth noting that
despite already erupting during last X7.1 flare on October 1, a more twisted MFR still appeared in the same PIL region before next X9.0 flare 38
hours later. It indicates that the formation of MFR driven by the shearing motion between the PIL was continuously ongoing during that period, which
was repeatedly but briefly interrupted by the series of flares. At this point, we have roughly answered the first question proposed in the end of
Introduction. But it is still unknown to us why the free energy and MFRs built up so fast in the PIL region that the major X9.0 and X7.1 flares
could occur in the same region within just 38 hours.

To explore the key factor for the rapid free energy build-up in AR 13842, we first analyzed the general evolution of photospheric magnetic fields in
this AR (see Figure \ref{fig3} and the corresponding animation). The initial conjugated magnetic polarity pair of AR 13842 consists of the leading
negative sunspot in the southwest and the following positive sunspot in the northeast (see N$_{1}$ and P$_{1}$ in Figure \ref{fig3}(a1)).
Since 2024 September 30, magnetic bipoles successively emerged between N$_{1}$ and P$_{1}$, and then separated along the northeast-southwest direction.
Meanwhile, continue flux emergence (see n$_{1}$ and p$_{1}$ in Figure \ref{fig3}(a2)) also took place to the southwest of N$_{1}$, and then separated
along the northwest-southeast direction, perpendicular to those of emerging bipoles between N$_{1}$ and P$_{1}$. As shown in Figure \ref{fig3}(a3),
collisional shearing then occurred between the opposite nonconjugated polarities (N$_{1}$ and p$_{1}$), producing the highly-sheared flaring PIL
\citep{2019ApJ...871...67C}. We further calculated the magnetic flux and the flux emergence rate of the whole AR, as well as the local region around
PIL. It is revealed that the unsigned flux of the whole AR gradually increased to reach $2.3\times10^{21}$ Mx late on October 3 with a maximum
emergence rate of $3.5\times10^{19}$ Mx hr$^{-1}$ (Figure \ref{fig3}(b1)).

Focusing on the flaring PIL region (Figures \ref{fig3}(b2)), its negative flux increased continuously to reach $8.7\times10^{20}$ Mx late on
October 3, which could be attributed to the persistent flux emergence of N$_{1-4}$. The positive magnetic flux in this region reached its
maximum ($2.5\times10^{20}$ Mx) around 22:20 UT on October 1, and then exhibited a significant and permanent decrease after each large flares,
accompanied by a slow overall decline. It is obvious that the abrupt drop of positive flux of p$_{1}$ is permanent magnetic change around
the flaring PIL for both the X7.1 and X9.0 flares near the disk center, suggesting that the photospheric magnetic field at the flaring PILs changes
from a more vertical to a more horizontal configuration after flares \citep{2001ApJ...550L.105K,2010ApJ...724.1218P,2012SoPh..277...59F,2018ApJ...852...25C,2019LRSP...16....3T}.
It is further supported by the rapid and irreversible increase of the horizontal magnetic field at the flaring PIL shown in Figure \ref{fig6}(b).

\begin{figure*}
\centering
\includegraphics [width=0.99\textwidth]{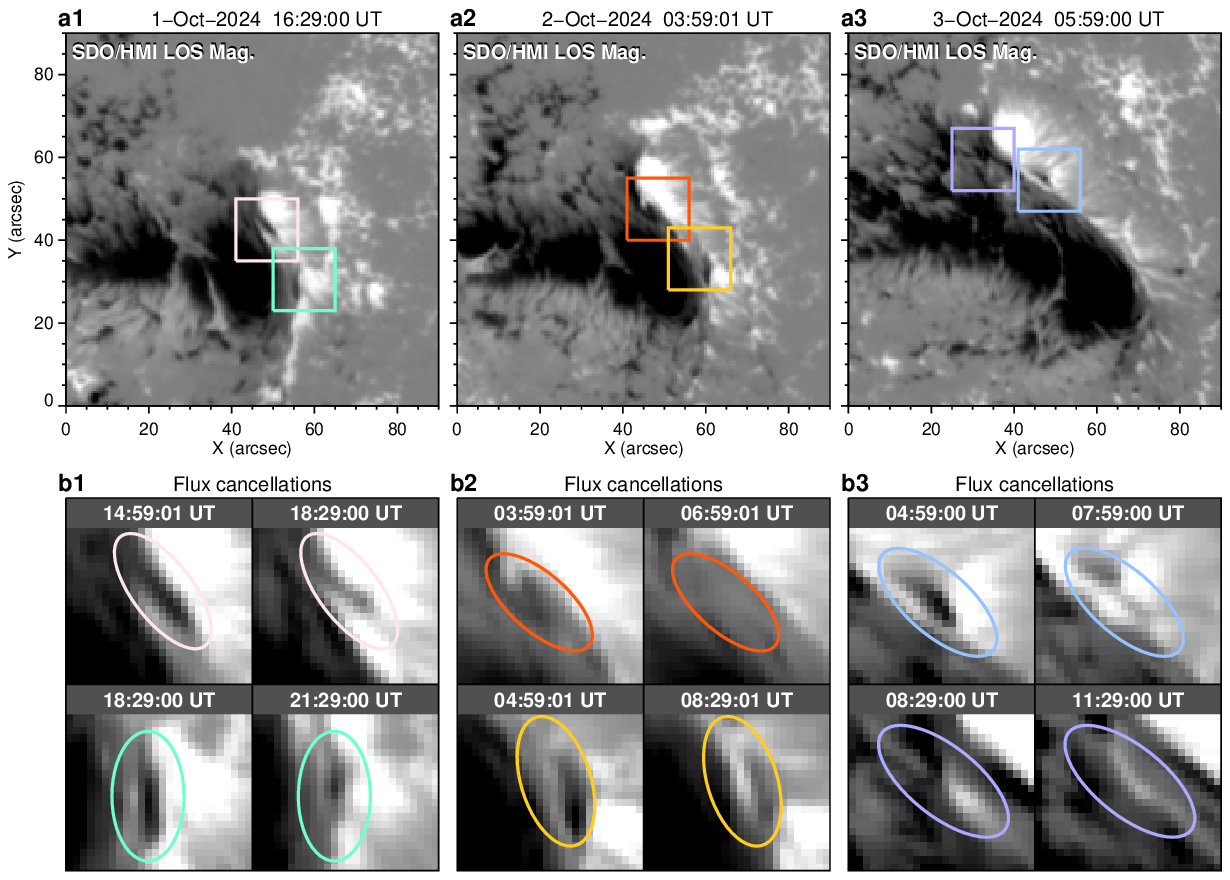}
\caption{Frequent magnetic flux cancellations occurring in the collisional shearing PIL region.
(a1)--(a3): Sequence of HMI LOS magnetograms exhibiting magnetic evolution of the PIL region during the phase of accelerated shearing after October 1.
Rectangles with different colors mark six typical flux cancellation events around the PIL region.
(b1)--(b3): Sequences of enlarged HMI LOS magnetograms exhibiting six flux cancellation processes. The colored ellipses denote the sites of the flux
cancellation.
}
\label{fig4}
\end{figure*}

As for the general positive flux decline, it is likely caused by persistent magnetic flux cancellation occurring between
nonconjugated N$_{1}$ and p$_{1}$ in the collisional PIL as investigated in \citet{2019ApJ...871...67C}. Through checking the HMI LOS magnetograms
of the PIL region, we indeed identified frequent flux cancellation events. Six typical flux cancellation processes are shown in Figure \ref{fig4}.
As reported by \citet{2023ApJ...955..105R}, pre-eruption MFRs are favored when magnetic cancellation occurs instead of simply due to shear
in the PIL region. We proposed that the frequent cancellations observed here facilitate repeated formations and eruptions of MFR above the PIL,
leading to multiple large flares in a relatively short time in AR 13842.

\begin{figure*}
\centering
\includegraphics [width=0.96\textwidth]{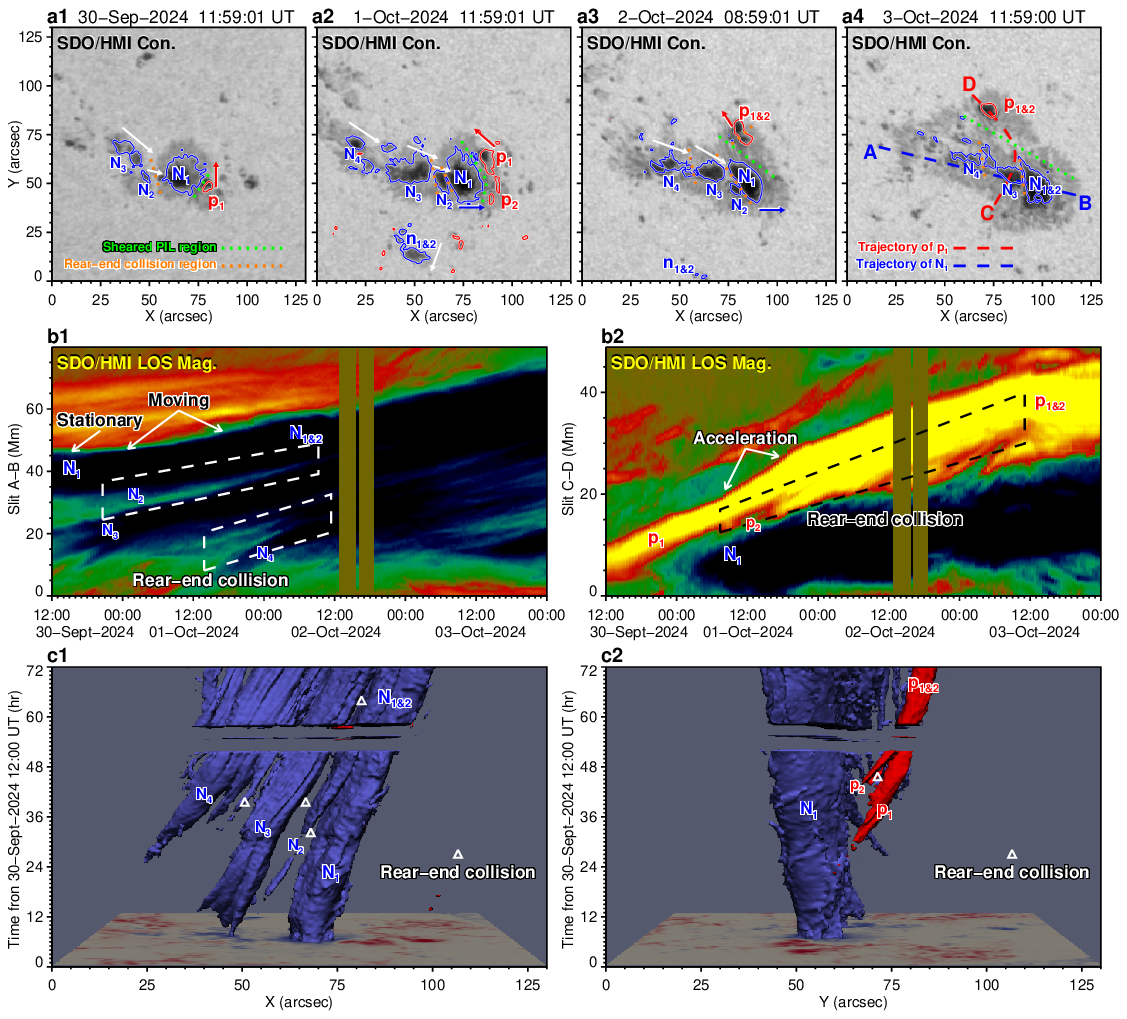}
\caption{Rear-end collisions between the same polarities on each side of a highly-sheared PIL in AR 13842.
(a1)--(a4): Sequence of HMI continuum intensity maps overlaid with the LOS magnetic field contours (red and blue curves) showing the evolution
of photospheric magnetic structures around the flaring sheared PIL. N$_{1}$ and p$_{1}$ label the main negative sunspot and the satellite
positive pore. N$_{2}$, N$_{3}$, N$_{4}$, and p$_{2}$ mark the magnetic patches emerging behind N$_{1}$ and p$_{1}$, respectively. Solid arrows
indicate the direction of motion of these photospheric magnetic structures.
(b1) and (b2): Time-distance maps derived from the HMI LOS magnetograms along the slits ``A--B" and ``C--D" shown in (a4), which respectively
approximate the trajectory of N$_{1}$ and p$_{1}$.
(c1) and (c2): 3D space-time representation of the rear-end collision between the same polarities using isosurfaces of magnetic field strength
as viewed from the south and the west. The spatial range of X and Y axes is the same as that of (a1)--(a4). The white triangles show the
collision sites.
}
\label{fig5}
\end{figure*}

Additionally, the magnetic patches on both sides of the flaring PIL region of AR 13842 also showed intriguing dynamic evolutionary features
driven by the vigorous flux emergence. As shown in Figures \ref{fig5}(a1)--(a4)), at the early evolution stage, the positive-polarity pore
(p$_{1}$) slowly drifted northward along the western side of the AR's main negative-polarity sunspot (N$_{1}$) while N$_{1}$ kept stationary.
Due to continuous magnetic flux emergence between the AR's two main sunspots, several negative magnetic patches (N$_{2}$, N$_{3}$, N$_{4}$)
successively appeared to the east of N$_{1}$ and then kept moving to N$_{1}$. Around October 1, N$_{1}$ was collided by these same-polarity
patches and then started to move westward (Figure \ref{fig5}(b1)). Here we name such interaction between the same polarities ``rear-end collisions"
on one side of a PIL. This dynamic evolution can be depicted well in a 3D representation by the time-stacking method (2D in space and 1D in
time, Figure \ref{fig5}(c1)). Similar rear-end collisions also occurred between p$_{1}$ and positive patches emerging behind (p$_{2}$) on the
other side of PIL and caused the acceleration of p$_{1}$ rotating around N$_{1}$ (Figures \ref{fig5}(a3), (b2), and (c2)).
As a result, a highly-sheared collisional PIL gradually formed between nonconjugated N$_{1}$ and p$_{1}$ from September 30 to October 3.
It's worth noting that the p$_{1}$ emerged synchronously (as part of a new flux emergence episode next to N$_{1}$) to the new emergence episode
in the pre-existing N$_{1}$-P$_{1}$ bipole in the east, which signifies ``Case A" and ``Case B" collisional shearing occurring in the same AR
(see Figure 15, \citet{2019ApJ...871...67C}).

\begin{figure*}
\centering
\includegraphics [width=0.8\textwidth]{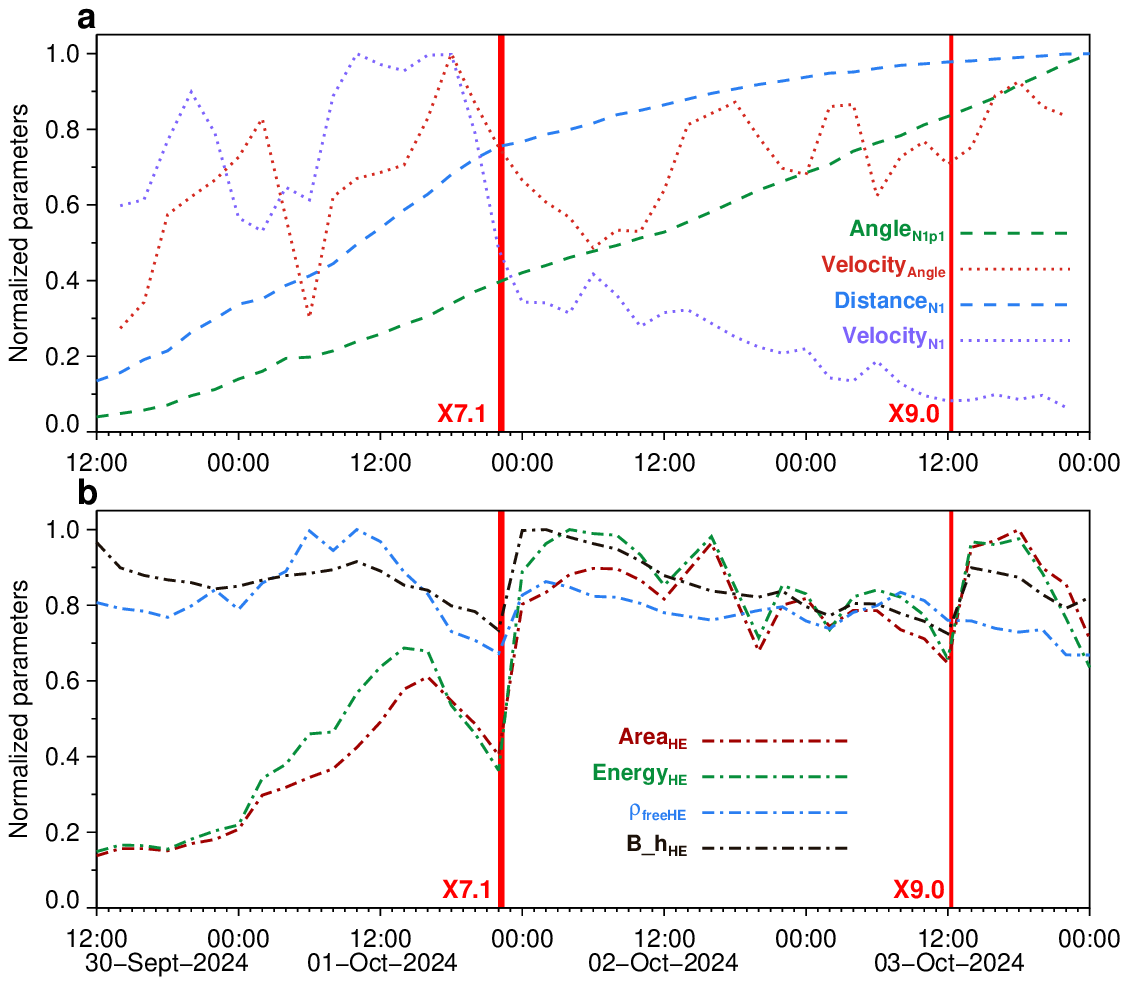}
\caption{Temporal profiles of kinematic parameters of the main sunspot (N$_{1}$) and the satellite pore (p$_{1}$), as well as the magnetic parameters
of the sheared PIL region with high $\rho_{free}$. The green, red, blue, and purple dashed/dotted curves in (a) show time evolutions of normalized rotation
angle of the line connecting intensity centroids of N$_{1}$ and p$_{1}$, the corresponding angle velocity, displacement distance of N$_{1}$, and the
corresponding movement speed, respectively. The maximum values of these kinematic parameters are 159.8$^\circ$, 3.8$^\circ$ hr$^{-1}$, 26.4 Mm, and
0.44 Mm hr$^{-1}$. The red, green, blue, and black dash-dotted curves in (b) show time evolutions of normalized area, total free energy, average
$\rho_{free}$, and average horizontal magnetic field strength of the PIL region with high $\rho_{free}$ ($\geq 9\times10^{4}$ erg cm$^{-3}$),
respectively. The maximum values of these magnetic parameters are $2.05\times10^{18}$ cm$^{2}$, $2.72\times10^{23}$ erg cm$^{-1}$,
$1.78\times10^{5}$ erg cm$^{-3}$, and 1544 G.
}
\label{fig6}
\end{figure*}

\begin{figure*}
\centering
\includegraphics [width=0.9\textwidth]{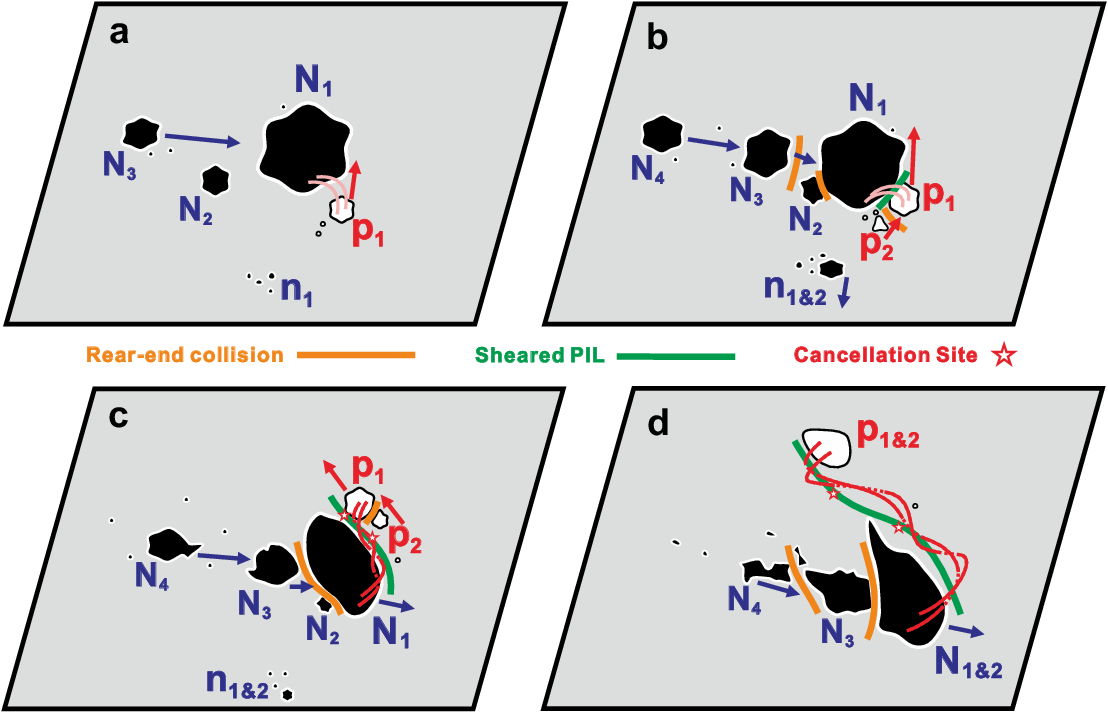}
\caption{Cartoon illustrating the rear-end collisions driven by persistent flux emergences, as well as the rapid formation of a collisional shearing PIL
and twisted MFR in AR 13842. The black patches represent the main negative sunspot (N$_{1}$) and a series of negative magnetic patches (N$_{2}$, N$_{3}$,
N$_{4}$) emerged to its east, as well as the conjugated negative polarity (n$_{1}$) of satellite positive pore (p$_{1}$). The white patches represent p$_{1}$
and a series of positive magnetic patches (e.g., p$_{2}$) emerged behind. The solid arrows indicate the direction of their motions. The yellow and green
ribbons mark the rear-end collision region between the same-polarity patches and collisional shearing PIL region between the opposite-polarity patches,
respectively. The red stars denote the flux cancellation sites around the PIL, and the pink curves and red twisted curves represent the sheared arcades
and MFRs above the PIL region.
}
\label{fig7}
\end{figure*}

In Figure \ref{fig6}, we quantitatively analyzed the kinematic evolution of N$_{1}$ and p$_{1}$ and photospheric magnetic parameters of the sheared
PIL region with high $\rho_{free}$. One can see that both of the movement speed of N$_{1}$ (purple dotted curve in Figure \ref{fig6}(a)) and rotation
angle velocity of the line connecting centroids of N$_{1}$ and p$_{1}$ (red dotted curve in Figure \ref{fig6}(a)) significantly increased later on
September 30, but the former one is several hours ahead. Almost synchronously, the area of the sheared PIL region with high $\rho_{free}$ and its
total free energy showed remarkable increase from the beginning of October 1 (see the red and green dash-dotted curves in Figure \ref{fig6}(b)).
These results support that when N$_{1}$ started to move after rear-ended by the successively emerging same-polarity patches, the shearing motion
between N$_{1}$ and p$_{1}$ was accelerated, which then promote the rapid accumulation of magnetic free energy in the sheared PIL region.

We also noticed another interesting fact in Figure \ref{fig6}(b) that the area and total free energy of the PIL region with high $\rho_{free}$
underwent a gradual decrease before the flare and a rapid increase after the flare. The rapid/irreversible enhancement of photospheric horizontal
magnetic field at the flaring PIL after flares have already been intensively investigated as a result of magnetic fields collapse (or contraction)
beneath coronal reconnection sites \citep{2012ApJ...745L...4L,2012ApJ...748...77S,2015RAA....15..145W,2019LRSP...16....3T}. As for the gradual
decrease of photospheric nonpotentiality prior to flares, \citet{2023ApJ...942...27L} recently found a long-term trend of decreasing twist in
the near-PIL regions before eruptive flares and proposed that when the pre-flare MFR lying above PIL slowly ascends, the area of photosphere cuts
through the MFR’s apex with large twist would be smaller and smaller, making the underlying photospheric fields less twisted and less horizontal.
As shown in Figure \ref{fig6}(b), the average horizontal magnetic field strength and free energy density in the PIL region studied here indeed
showed a gradual decrease before the flare. Combining the results shown in Figure \ref{fig2}, we believe that before the X7.1 and X9.0 flares in
AR 13842, both twisted MFRs had already been formed above the PIL driven by the shearing motion and flux cancellations and then gradually
ascended \citep{2005ApJ...630.1148S,2012ApJ...756...59L}, which could be responsible for the gradual decrease of photospheric nonpotentiality
prior to flares observed here.

To illustrate more intuitively the rear-end collisions driven by persistent flux emergences, as well as the rapid formation of a collisional
shearing PIL and twisted MFR in AR 13842, we sketch a cartoon model in Figure \ref{fig7}. When the satellite positive pore (p$_{1}$) approached
the main negative sunspot (N$_{1}$), connection between the two regions with opposite polarities would be then formed through flux cancellation
during the following shearing motion. Accelerated by the rear-end collisions occurring on N$_{1}$ (collided by N$_{2}$, N$_{3}$ and N$_{4}$) and
p$_{1}$ (collided by p$_{2}$), a highly-sheared collisional PIL rapidly formed between nonconjugated N$_{1}$ and p$_{1}$, where a great deal of
free energy was accumulated and the magnetic fields above also became more sheared arcades.
As illustrated by the tether-cutting model \citep{2001ApJ...552..833M}, such shearing motion accompanied by flux cancellation is crucial for the
formation of MFR. Moreover, when drifting around N$_{1}$ in a counterclockwise direction, p$_{1}$ itself also slowly rotated clockwise (see the
animation associated with Figure \ref{fig3}), which can also facilitate the formation of a twisted MFR \citep{2015ApJS..219...17Y,2017ApJ...849L..21Y}.

\section{Summary}\label{sect6}
Based on the \emph{SDO} and \emph{CHASE} observations, we investigate the formation of a highly-sheared PIL region in AR 13842, which
produced the largest X9.0 flare of Solar Cycle 25 so far and another X7.1 flare within two days. The main results are summarized as follows:
\begin{enumerate}
\item AR 13842 produced two X-class (X7.1 and X9.0) and around ten M-class flares within just 3 days from 2024 October 1 to October 3.
Both of the X7.1 and X9.0 flares had pre-flare hot EUV sigmoidal structures and twisted MFRs above the same sheared PIL and produced multiple
flare ribbons, including two main ribbons observed in white-light wavelength.

\item AR 13842 initially consisted of a conjugated magnetic polarity pair (N$_{1}$ and P$_{1}$). Magnetic bipoles then successively emerged
between N$_{1}$ and P$_{1}$, and separated along the northeast-southwest direction. Meanwhile, continue flux emergence (n$_{1}$ and p$_{1}$)
also took place to the southwest of N$_{1}$, and then separated along the northwest-southeast direction. As a result, collisional shearing
occurred between the opposite nonconjugated polarities (N$_{1}$ and p$_{1}$), accompanied by frequent flux cancellations. And a highly-sheared
flaring PIL began to form.

\item At the early evolution stage of the flare-productive PIL, the satellite positive pore (p$_{1}$) approached and then slowing drifted
northward along the western side of N$_{1}$. N$_{1}$ kept stationary until a series of negative magnetic patches emerged and continuously
approached it. Rear-ended by these patches with the same polarity, N$_{1}$ started moving westward around October 1. Similar rear-end
collisions also occurred between p$_{1}$ and positive patches emerging behind on the PIL's other side. Such rear-end collisions driven by
persistent flux emergences on both sides of PIL dramatically accelerated the collisional shearing motion between nonconjugated N$_{1}$ and p$_{1}$.

\item During the rapid formation of the collisional shearing PIL, a great deal of free magnetic energy was also synchronously accumulated. And twisted
MFRs were also rapidly formed driven by the shearing motion and frequent flux cancellations in the PIL, where multiple major flares successively
occurred. Before the large flares, the area and total free energy of the high-free-energy-density PIL region gradually decreased in the photosphere,
which could indicate the slow ascent of MFR before eruption and serve as a precursor of solar eruptions.
\end{enumerate}

\acknowledgments
The authors appreciate the anonymous referee for the constructive comments and valuable suggestions.
The data used here are courtesy of the \emph{SDO}, \emph{CHASE}, and \emph{GOES} science teams. \emph{SDO} is a mission of NASA's
Living With a Star Program. \emph{CHASE} mission is supported by CNSA. Y.J.H. appreciates Dr. Huidong Hu and Qiao Song for their
helpful suggestions. The authors are supported by the Strategic Priority Research Program of CAS (XDB0560000), the National Natural
Science Foundation of China (12273060, 12333009, 12573056, and 12533010), the National Key R\&D Program of China (2022YFF0503800),
the Fundamental Research Funds for the Central Universities (KG202506), the Youth Innovation Promotion Association CAS (2023063),
the Youth Research Special Project of NCUT (2025NCUTYRSP038), Research Startup Fund of NCUT (11005136025XN076-072), China's Space
Origins Exploration Program (GJ11020405), and the Specialized Research Fund for State Key Laboratory of Solar Activity and Space Weather.

\bibliography{ref}{}
\bibliographystyle{aasjournal}

\end{document}